\documentclass[%
 aip,
 amsmath,amssymb,
reprint,
]{revtex4-1}

\usepackage{graphicx}
\usepackage{dcolumn}
\usepackage{bm}

\usepackage[final]{hyperref} 
\hypersetup{
	colorlinks=true,       
	linkcolor=blue,        
	citecolor=blue,        
	filecolor=magenta,     
	urlcolor=blue         
}
\usepackage{siunitx}
\usepackage{upgreek}
\usepackage{tabularx} 
\usepackage[utf8]{inputenc}
\usepackage[T1]{fontenc}
\usepackage{mathptmx}
\usepackage{etoolbox}

\usepackage{xcolor}
\usepackage{xspace}
\usepackage[final]{changes}
\setdeletedmarkup{\color{red}\sout{#1}}
\definechangesauthor[color=blue   , name={Brian D'Urso}]{BD}
\definechangesauthor[color=green   , name={Connor Murphy}]{CM}

\begin{document}

\preprint{}

\title[Selective Loading of a Micrometer-Scale Particle into a Magneto-Gravitational Trap by Sublimation-Activated Release]{Selective Loading of a Micrometer-Scale Particle into a Magneto-Gravitational Trap by Sublimation-Activated Release}
\author{Connor E. Murphy}
\author{Mario Duenas}
\author{Daniel Iron}
\author{Tobias Nelson}
\author{Brian D'Urso}
\email{durso@montana.edu}
\affiliation{Department of Physics, Montana State University, Bozeman, Montana 59717, USA}
\date{\today}

\begin{abstract}
In this paper we discuss a technique for selectively loading a particle into a magneto-gravitational trap using the sublimation of camphor to release particles from a tungsten probe tip directly into the trapping region. This sublimation-activated release (SAR) loading technique makes use of micropositioners with tungsten probe tips, as well as the relatively fast rate of sublimation of camphor at room temperature, to selectively load particles having diameters ranging from $\SI{8}{\micro \meter}$ to $\SI{100}{\micro \meter}$ or more. The advantages of this method include its ability to selectively load unique particles or particles in limited supply, its low loss compared to alternative techniques, the low speed of the particle when released, and the versatility of its design which allows for loading into traps with complex geometries. SAR is demonstrated here by loading a particle into a magneto-gravitational trap, but the technique could also be applicable to other levitated optomechanical systems.
\end{abstract}

\maketitle

\section{Introduction}
\label{sec:intro}

Levitated optomechanical systems -- in which a typically small test mass or particle is held in place by a trapping potential, often placed under vacuum, and tracked by measuring light scattered off of it from an illumination source -- are widely used to create well-isolated environments for performing sensitive measurements. In particular, such systems can be highly sensitive to small forces \cite{liang2023yoctonewton,rodenburg2016quantum,ranjit2015attonewton,ranjit2016zeptonewton,merenda2009three}, making it possible to perform experiments capable of 
precisely measuring gravitational effects \cite{lewandowski2020towards}, searching for dark matter interactions\cite{afek2022coherent,brady2023entanglement,higgins2024maglev}, detecting neutrinos\cite{carney2023searches}, or testing the limits of quantum mechanics \cite{bateman2014near,neumeier2024fast,millen2016cooling}.
The wide range of possible particle sizes and trapping potentials offers a vast parameter space for exploring a variety of physical phenomena.

In addition to allowing for high-sensitivity force measurements, these systems can also be used to isolate and study the properties of individual particles of exotic materials such as single particle reactors in the atmosphere \cite{ai2023optically}, graphene flakes \cite{kane2010levitated}, \added{micromagnets\cite{wang2019dynamics,gieseler2020single}, superconducting microspheres\cite{romero2012quantum,hofer2023high}}, or nano-diamonds containing NV color centers \cite{kuhlicke2014nitrogen,delord2017electron,hsu2016cooling}.

Commonly used types of traps are optical traps, which create a stable potential for trapping dielectric particles using a tightly focused laser beam\cite{ashkin1971optical,summers2008trapping,li2010measurement,gieseler2012subkelvin,asenbaum2013cavity,mestres2015cooling,millen2016cooling,rondin2017direct,khodaee2022dry,neumeier2024fast,schmidt2012metrology,park2017optical,grass2016optical,ashkin1987optical,schmidt2012reconfigurable,liang2023yoctonewton,lindner2024hollow}; Paul traps, which use alternating electric fields to trap charged particles \cite{bykov2019direct,delord2017electron,kane2010levitated,conangla2018motion,kuhlicke2014nitrogen}; magneto-gravitational traps (MGTs), which form trapping potentials for diamagnetic particles from a combination of Earth's gravitational field and static magnetic fields \cite{doi:10.1063/1.5051667,slezak2018cooling,hsu2016cooling,lewandowski2020towards}; and hybrid traps \cite{millen2015cavity}, which combine, e.g., optical and Paul traps to more precisely manipulate the trapping potential.

\subsection{Alternative Loading Techniques}
\label{subsec:problem}

One challenge that is common to all levitated optomechanical systems is that of placing a suitable test mass into the trapping potential, or ``loading'' a particle into the trap.

The origin of the challenge is that micrometer and sub-micrometer scale particles may be held to surfaces by a combination of van der Waals, electrostatic, and capillary forces which can be much larger than their weight\cite{zhou1998adhesion,chen2009overcoming,ashkin1971optical}, making them easy to pick up but difficult to drop in a controlled fashion. Most loading techniques rely on giving the particle a mechanical kick to release it from its loading surface. Table~\ref{tab:loading} shows a comparison of several common techniques and some selected parameters for comparison between them.

\begin{table*}
    \begin{center}
    \caption{Comparison of Loading Techniques}
    \begin{tabularx}{\textwidth}{| >{\raggedright\arraybackslash}X | >{\centering\arraybackslash}X | >{\centering\arraybackslash}X | >{\centering\arraybackslash}X | >{\centering\arraybackslash}X | >{\centering\arraybackslash}X |}
        \hline
        Method & Wet/Dry & Pressure\footnote{For this section, Atm. refers to atmospheric pressure (760 Torr), R.V. refers to pressures between atmospheric pressure and $\sim 10^{-3}$ Torr, H.V. to pressures between $\sim 10^{-3}$ Torr and $\sim 10^{-8}$ Torr, and U.H.V. for pressures lower than $\sim 10^{-8}$ Torr.} & Selective & Non-Contaminating & Size\footnote{\added{Refers to the sizes reported in literature, not necessarily the actual limits on the capabilities of the loading technique.}} \\
        \hline
        \hline
        SAR (this paper) & Dry & Atm. & Yes & Yes & >\SI{8}{\micro \meter} \\
        \hline
        Piezo        \cite{li2010measurement,lewandowski2020towards,khodaee2022dry,hsu2016cooling,slezak2018cooling}& Dry & Atm.-R.V. & No & No & >\SI{43}{\nano \meter} \\
        \hline
        LIAD \cite{kuhn2015cavity,asenbaum2013cavity,millen2016cooling,bykov2019direct} & Dry & R.V.-\added{U.}H.V. & No & No & 100-\SI{1000}{\nano \meter} \\
        \hline
        Load-Lock \cite{mestres2015cooling} & Dry & H.V. & No & Yes & \SI{75}{\nano \meter} \\
        \hline
        Hollow Core \cite{schmidt2012reconfigurable,grass2016optical,schmidt2012metrology,lindner2024hollow} & Dry & Atm.-U.H.V. & No & Yes & 100-\SI{755}{\nano \meter} \\
        \hline
        Piezo/Optical \cite{ashkin1971optical,park2017optical} & Dry & Atm. & Yes & Yes & \added{15-\SI{25}{\micro \meter}} \\
        \hline
        Electrospray \cite{kuhlicke2014nitrogen,conangla2018motion,kane2010levitated} & Wet & Atm. & No & No & <\SI{1}{\micro \meter} \\
        \hline
        ``Tapping'' \cite{doi:10.1063/1.5051667,delord2017electron} & Dry & Atm. & No & No & >\SI{1}{\micro \meter} \\
        \hline
        Atomizer \cite{summers2008trapping,gieseler2012subkelvin,liang2023yoctonewton} & Wet & Atm. & No & No & \added{\SI{70}{\nano \meter}-\SI{3.01}{\micro \meter}} \\
        \hline
    \end{tabularx}
    \label{tab:loading}
    \end{center}
\end{table*}

There are several available methods for loading particles into a trap, both under atmospheric pressure and directly into a chamber under vacuum.

When loading into a chamber at atmospheric pressure, air resistance typically provides enough damping to reduce the energy of the particle so that it can be trapped in a conservative potential. One of the simpler loading techniques available is to tap an object such as a wire\cite{delord2017electron} or a cotton swab\cite{doi:10.1063/1.5051667} to knock particles off the surface and into the trapping region. Similarly, the piezo loading method\cite{lewandowski2020towards,khodaee2022dry,hsu2016cooling,slezak2018cooling} involves picking up several particles on a surface, then using a voltage-driven piezoelectric actuator to vibrate that surface and knock those particles off the surface. Another method commonly used is to spray a suspension of particles in a liquid using a process known as electrospray ionization\cite{mcluckey1991ion,kuhlicke2014nitrogen,conangla2018motion,kane2010levitated}. After being trapped, the liquid portion of the suspension evaporates, leaving behind trapped particles. An atomizer can be used to achieve a similar effect\cite{summers2008trapping,gieseler2012subkelvin,liang2023yoctonewton}. In general, all of these techniques release particles with high velocities at uncontrolled angles. A hybrid piezo-optical approach using a vibrating surface to release particles and a laser aimed through the surface to push a single particle into a trap improves the selectivity of loading~\cite{ashkin1971optical,park2017optical} relative to piezo-only approaches.

To load particles into a chamber under vacuum, different methods can be used to reduce the kinetic energy of the particle enough for it to be trapped: the trap potential can be altered as the particle is being loaded\cite{bykov2019direct}, optical cooling can be used to provide the necessary damping provided that the launch energy is low enough\cite{millen2016cooling,asenbaum2013cavity}, or the particle can be carefully placed into position by some kind of transfer method\cite{grass2016optical,lindner2024hollow,schmidt2012reconfigurable,mestres2015cooling,schmidt2012metrology}. One method for directly loading into vacuum is laser-induced acoustic desorption (LIAD)\cite{golovlev1997laser,kuhn2015cavity,sezer2015laser,asenbaum2013cavity,millen2016cooling,bykov2019direct}, which releases particles from a surface by striking the back side of it with a high-energy pulse of laser. When combined with a technique to decrease the kinetic energy of the particle as it is being loaded, this allows particles to be loaded directly into a trap under vacuum.

In each of the loading methods described above, with the exception of the piezo-optical hybrid technique, many particles are sprayed into the trap at once, leading to potential contamination of trap components and the vacuum chamber in which they are housed. The load-lock loading method\cite{mestres2015cooling}, in which a particle is loaded using a contaminating method and then transferred to a chamber to be loaded a second time in a non-contaminating fashion, can be used to keep the main test chamber free of unwanted particles. Another method for doing this is to use one of the contaminating loading methods to load a particle into a hollow-core fiber, then use an optical conveyor belt to transfer that particle into a clean chamber\cite{grass2016optical,lindner2024hollow,schmidt2012reconfigurable,schmidt2012metrology}. Both of these methods allow for loading directly into vacuum, but neither allows particles to be individually selected or to be loaded efficiently without losing many other particles in the process.

Many available methods for loading particles into a trap are highly inefficient, non-selective, and risk contaminating the chamber, optics, and trap components with a large number of untrapped particles. This may be acceptable when trapping nearly identical and easily obtained particles such as silica nano- or micro-spheres, but may not be ideal when considering irregularly shaped particles or particles in a limited supply such as graphene flakes or diamonds containing NV color centers. For such applications, it would be beneficial to have a way to select one specific particle (e.g. after external characterization) and load it with a low rate of loss.

In this paper, we introduce a new sublimation-activated release (SAR) loading technique that is selective, non-contaminating, and could be usable with a variety of trap and vacuum-chamber geometries. SAR loading works by coating a loading surface in a material that sublimates at room temperature (such as camphor) before picking up a selected particle. The loading surface (with the particle attached) is then placed in the trapping region. When the camphor sublimates, the particle is released with a low velocity into the trap. We demonstrate the SAR technique by loading \SI{8}{\micro\meter} silica microspheres into a MGT housed in a vacuum chamber at atmospheric pressure using a camphor-coated tungsten probe tip mounted on a micropositioner.

\subsection{Magneto-Gravitational Traps}
\label{subsec:bg}

MGTs make use of static magnetic fields to trap diamagnetic particles, in contrast to optical traps and Paul traps which use oscillating electromagnetic fields to trap particles. The static fields used in a MGT make it simple to pump the chamber down to ultra-high vacuum (UHV) while keeping a particle trapped without excessive heating of the particle or its motion\cite{hsu2016cooling}.

\begin{figure}
\centering \includegraphics[width=1.0\columnwidth]{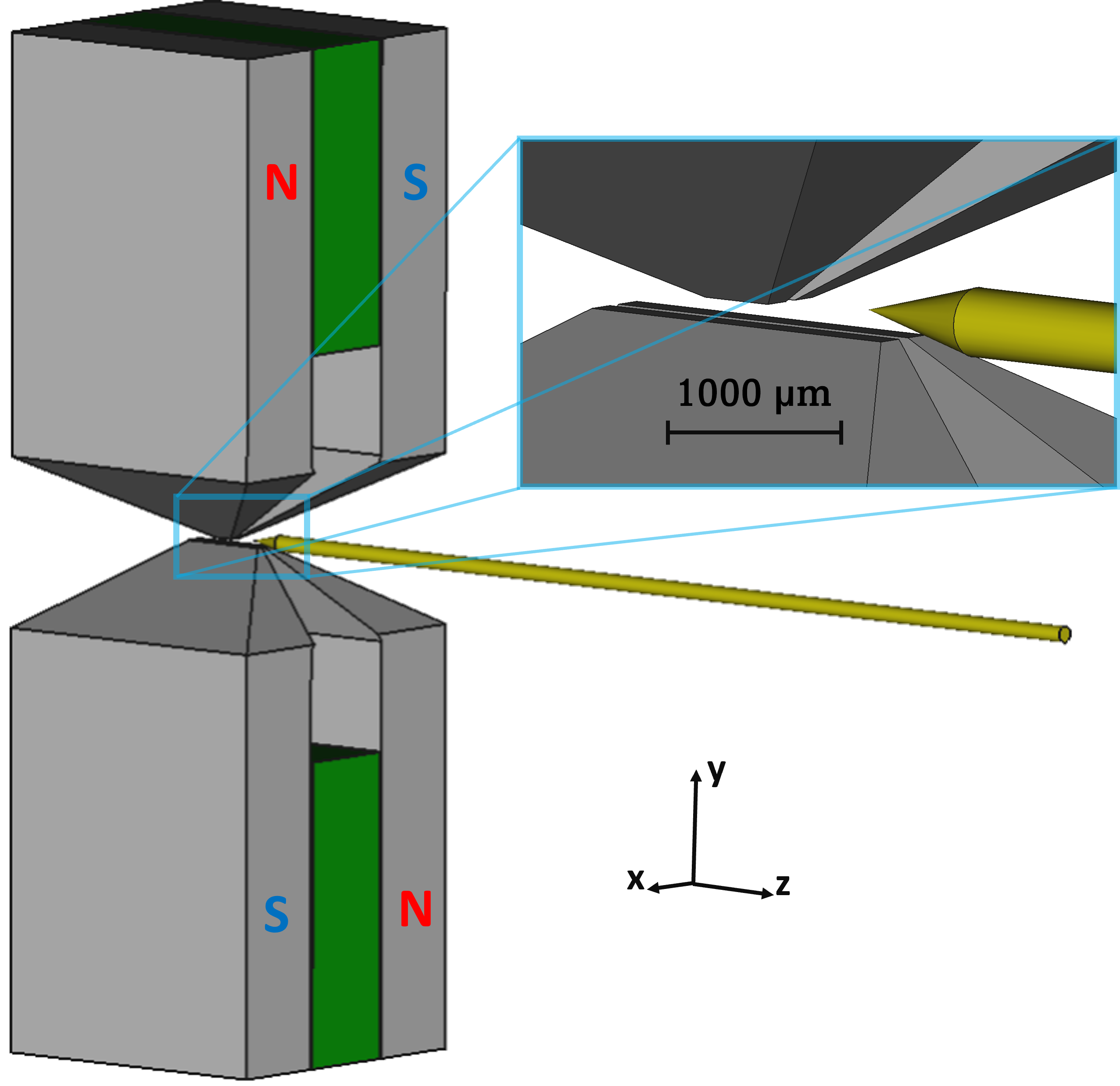}
\caption{\label{fig:trap} A diagram of the MGT used to develop the loading technique. The four pole pieces produce a quadrupole field in the vertical (y) and transverse (x) direction, and the asymmetry between the bottom set and the top set of pole pieces produces a higher-order effect that works with gravity to form a trapping potential in the axial (z) direction. A model of the tungsten probe used for loading is shown to provide an approximate size comparison. The probe is removed after loading.}
\end{figure}

The procedure described in this paper was developed for a MGT having four magnetic poles generated by four Hiperco 50A (Ed Fagan, Inc) pole pieces and two permanent magnets, as shown in Figure~\ref{fig:trap}. The trapping potential can be described by\cite{slezak2018cooling}

\begin{equation} \label{eq:potential}
U = mgy - \frac{\chi V}{2 \mu_0} B^2,
\end{equation}

\noindent
where $m$ is the mass of the particle, $\chi$ is its volume magnetic susceptibility, $V$ is its volume, and $\vec{B}$ is the trap magnetic field.

This trap design allows the trapping of diamagnetic particles with diameters from \SI{1}{\micro \meter}\cite{slezak2018cooling} to \SI{100}{\micro \meter} or more \cite{10.1117/12.2515721}. The trap is mounted in a vacuum chamber, and particles are loaded from the side through an open port on the chamber at atmospheric pressure.

\section{Trap Loading via Sublimation-Activated Release}
\label{sec:mcl}

We demonstrate that SAR allows a selected particle to be loaded into a MGT with a high probability of success. The technique has been used with diamagnetic particles ranging from approximately $\SI{8}{\micro \meter}$ to $\SI{100}{\micro \meter}$ in diameter.

The critical step in this technique is coating the loading surface in a suitably sticky substance that sublimates at room temperature, allowing the attached particle to fall into the trap in a controlled fashion.

Experimentally, we have found that camphor is a suitable choice of coating material due to its low cost, low toxicity, waxy texture, and moderate room-temperature sublimation rate.

We used tungsten probe tips mounted on micropositioners for our loading surface, and silica microspheres to demonstrate the technique. In more detail, the steps to select and load a single particle include:

\begin{enumerate}
    \item{A loading apparatus is made by mounting a tungsten probe (Singatone SE-20T) on a 10 cm shaft (Signatone S3-E-1) and connecting that to a Signatone S-931 micropositioner. Figure~\ref{fig:apparatus} shows the \added{trapping} setup we used for the results presented in this paper.}

\begin{figure}[ht!] 
\centering \includegraphics[width=1.0\columnwidth]{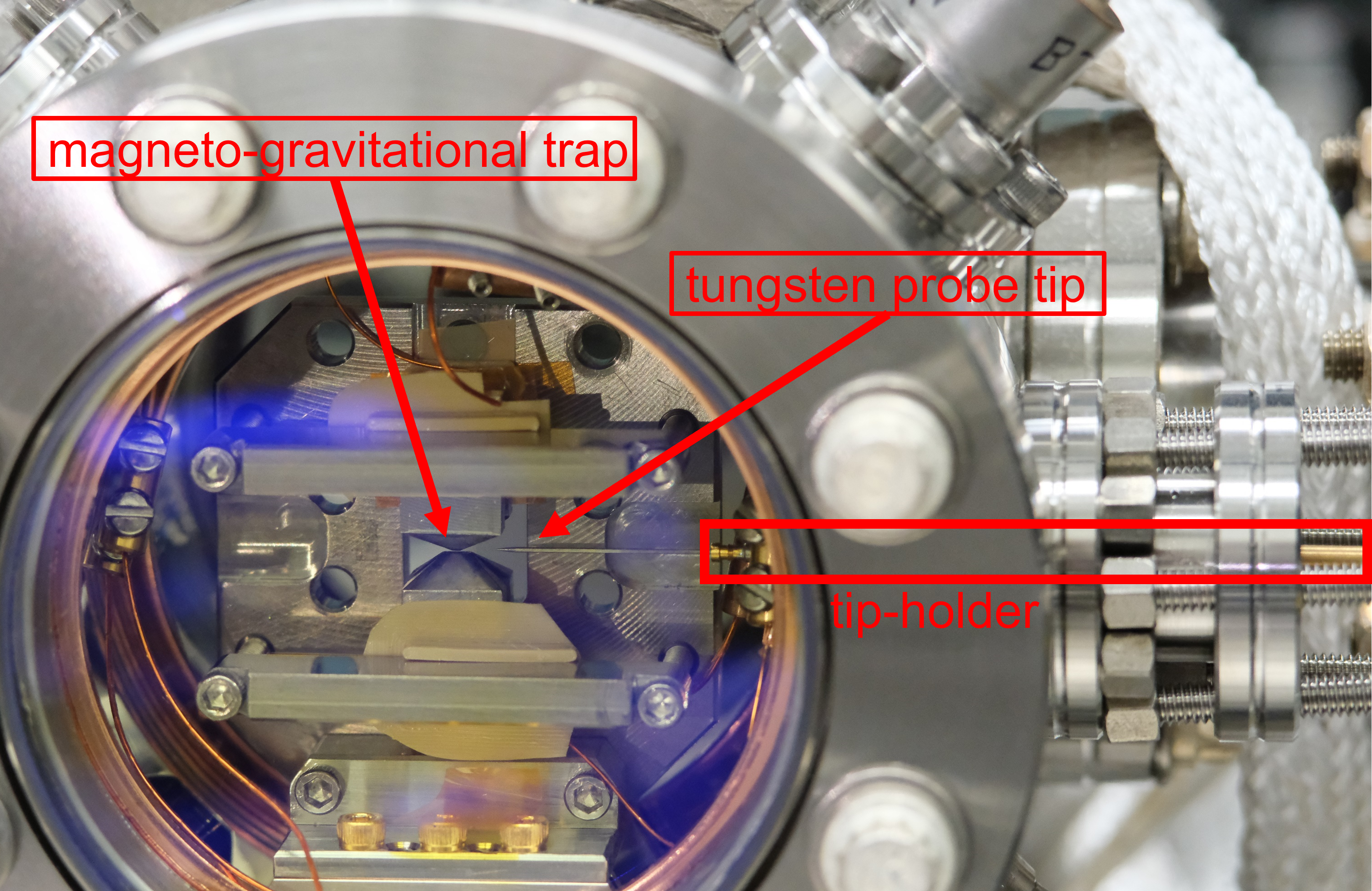}
\caption{\label{fig:apparatus} Image of the apparatus used for loading particles into a MGT via the SAR loading technique. The MGT is shown housed in a vacuum chamber, with the tungsten probe tip used for loading positioned next to it. The tip-holder shown is a \SI{10}{\centi\meter} rod attached on its other end to a micropositioner (not shown). \added{Before being positioned in the trap as shown here, the tip is coated with camphor and receives a selected particle for loading as in Figs.~\ref{fig:slide} and \ref{fig:steps}.}}
\end{figure}

    \item{The particles are placed on an aluminum-foil covered slide and spread out by tapping the edge of the slide. Next, an ionizing radiation source (e.g. Am-241) is brought near the particles to decrease excess charge on the particles\added{, as shown in Figure~\ref{fig:slide}}. Without this step, the particles tend to stick to the camphor without being released.}

\begin{figure}[ht!] 
\centering \includegraphics[width=1.0\columnwidth]{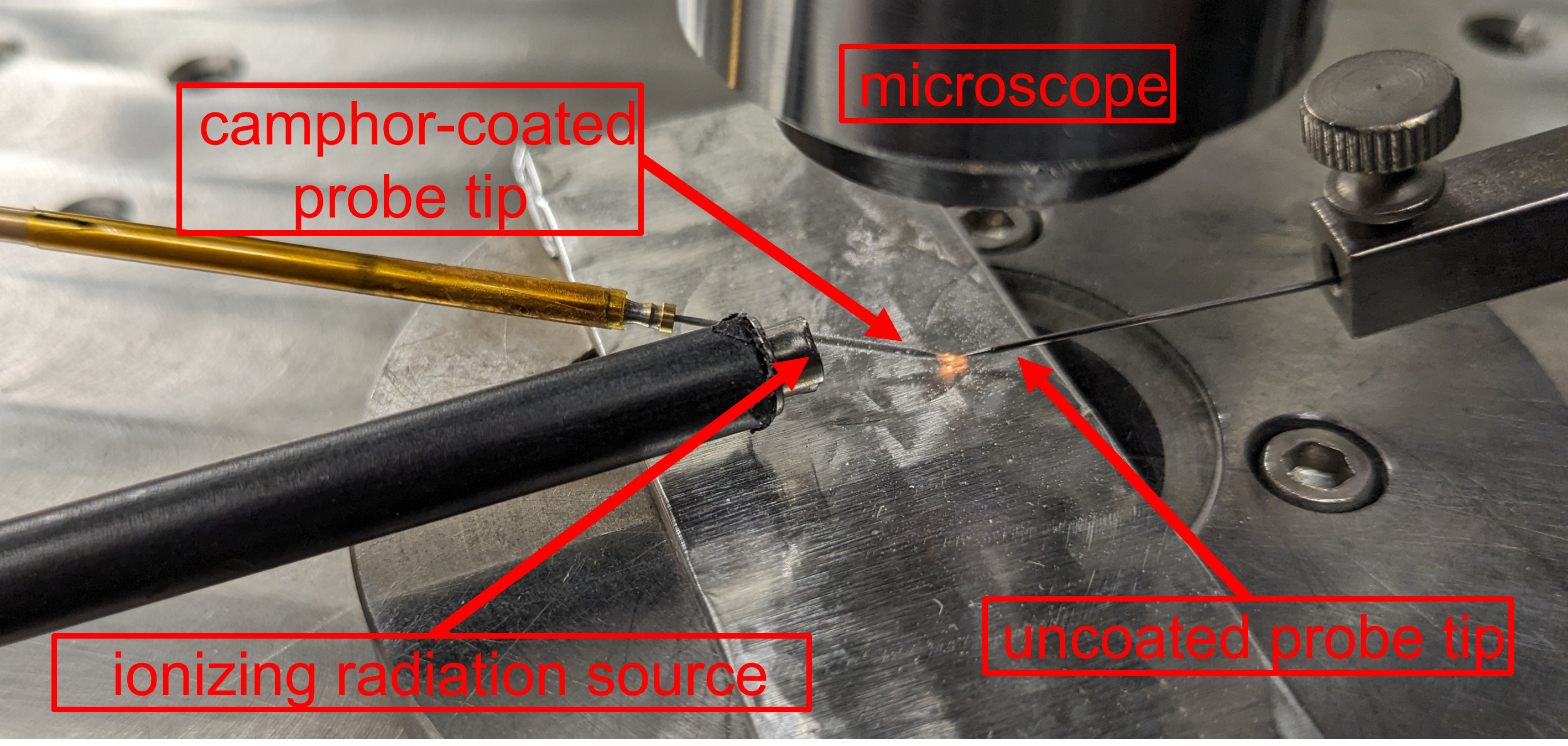}
\caption{\label{fig:slide} \added{Image of the setup for selecting a particle to be loaded from an aluminum-foil covered slide. The desired particle is transferred from the slide to the camphor-coated probe tip using the second, uncoated probe tip (see Figure~\ref{fig:steps}). The radiation source is used to decrease the charge on the particles and the camphor.}}
\end{figure}

    \item{Approximately $\SI{18}{g}$ of crushed camphor is placed at the bottom of a copper crucible covered by aluminum foil with a \SI{0.5}{\centi \meter} hole in the top and heated at $\SI{225}{\celsius}$ for at least 20 minutes to increase the concentration of camphor vapor in the crucible (Figure~\ref{fig:heating}). While heating, the hole is covered by a glass lid.}

\begin{figure}[ht!] 
\centering \includegraphics[width=1.0\columnwidth]{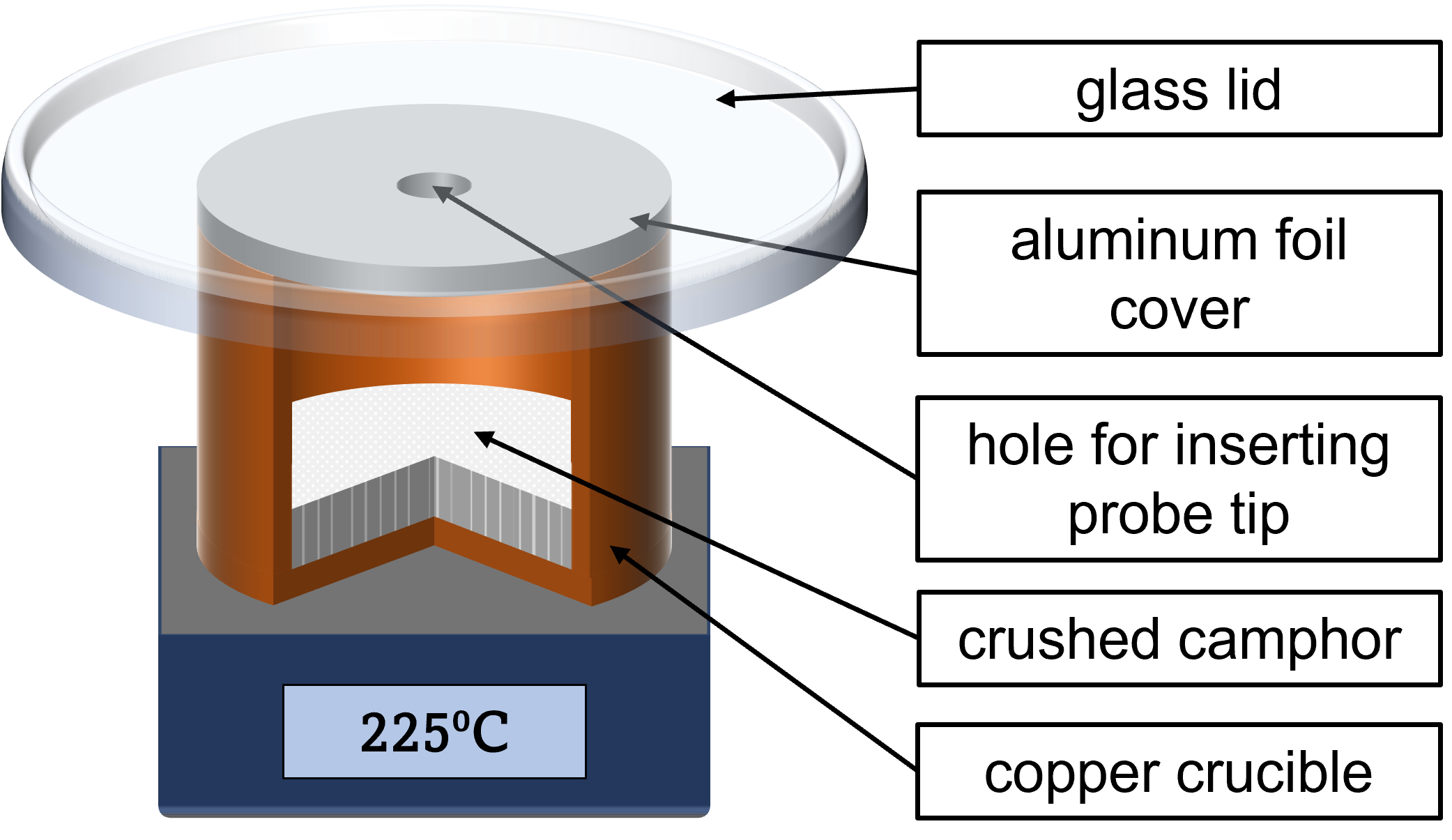}
\caption{\label{fig:heating} Diagram of the setup used to heat the camphor. A mortar and pestle were used to crush the camphor into a loose millimeter-scale grain powder which was then placed at the bottom of a copper crucible. The aluminum foil covering the crucible is folded over itself twice and has a hole in the top that is approximately \SI{0.5}{\centi \meter} in diameter. The hole is covered by a flat glass lid at all times except when the tungsten probe tip is being coated in order to conserve camphor. Over time the camphor escapes and builds up on the aluminum foil, so the camphor must be scraped off and replenished occasionally.}
\end{figure}

    \item{One particle is chosen under a microscope to be loaded into the trap by moving other particles away using a micropositioner-mounted probe, as necessary. At this point the particle is picked up by an uncoated tungsten probe tip. This prevents accidentally picking up unwanted particles, and it allows the particle to be pressed firmly into the camphor at a desired location on the camphor-coated tip as described in step 5.}
    \item{The tip of the loading apparatus is inserted into the hole in the aluminum foil for approximately 10 seconds. This usually results in large crystals forming on the tip, which are not ideal for precise positioning. 
    \replaced{In principle a faster rate of deposition of the camphor results in the formation of smaller crystal structures, but in practice we found that a smooth coating of camphor could be formed by allowing the large crystal structures to sublimate before attempting to pick up a particle.}{ At this point the camphor coating is monitored until enough of the crystals have sublimated to leave behind only a smooth coating of camphor.}
    The camphor is exposed to the ionizing radiation source during this time to decrease the net charge on the camphor. The particle selected in step 4 is then transferred to the camphor-coated tip (Figure~\ref{fig:steps}).}

\begin{figure}[ht!] 
\centering \includegraphics[width=1.0\columnwidth]{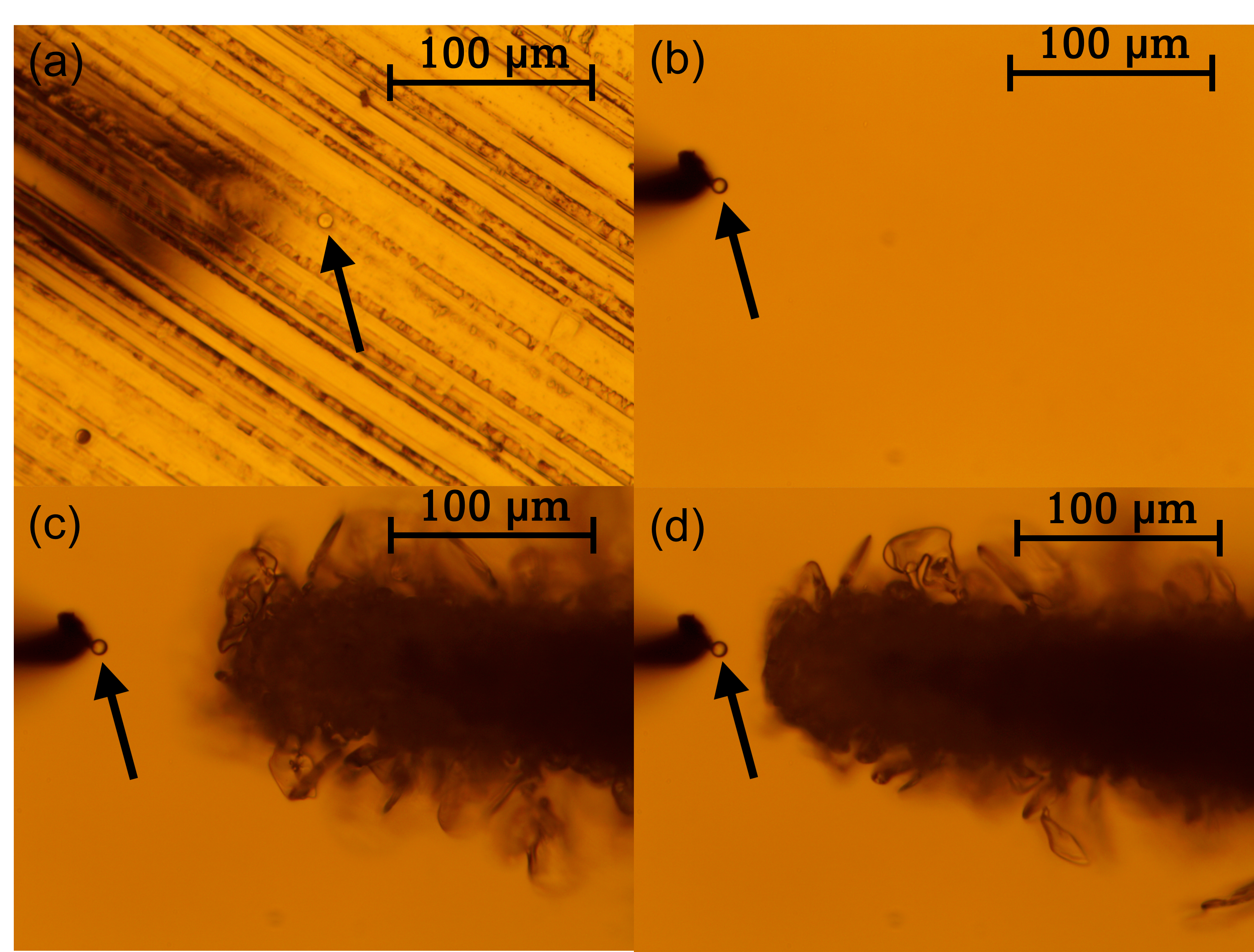}
\caption{\label{fig:steps} The process for loading an $\SI{8}{\micro \meter}$-diameter silica microsphere is shown above. The particle is selected from the sample set under a microscope, and an uncoated probe tip (the blurry shape to the left of the particle) is brought near above the foil (a). The particle is then picked up by the uncoated probe tip (b). A second probe tip is then coated in camphor for 10 seconds, but this often leads to the growth of unwanted large crystal structures (c). After waiting about 1-3 minutes the large crystals will have sublimated enough to allow the particle to be pressed into the camphor using the uncoated probe (d).}
\end{figure}

    \item{The probe apparatus is transferred to a swivel (or long translation stage) just outside the MGT and vacuum chamber, where the camphor-coated tip is inserted carefully through an open vacuum chamber port and into the trapping region between the pole pieces. This is done with the aid of a camera focused between the pole pieces that form the trap, back-illuminated by an LED. One of the most challenging parts of this process is making sure that the particle is in the trapping region when it is released. If the depth of field of the objective is on the same scale as the size of the trapping region, having the particle in focus may be enough precision for loading (Figure~\ref{fig:trapping}). Figure~\ref{fig:trapped} shows an overlay of several images taken at the time a particle was released into the trap.}

\begin{figure}[ht!] 
\centering \includegraphics[width=1.0\columnwidth]{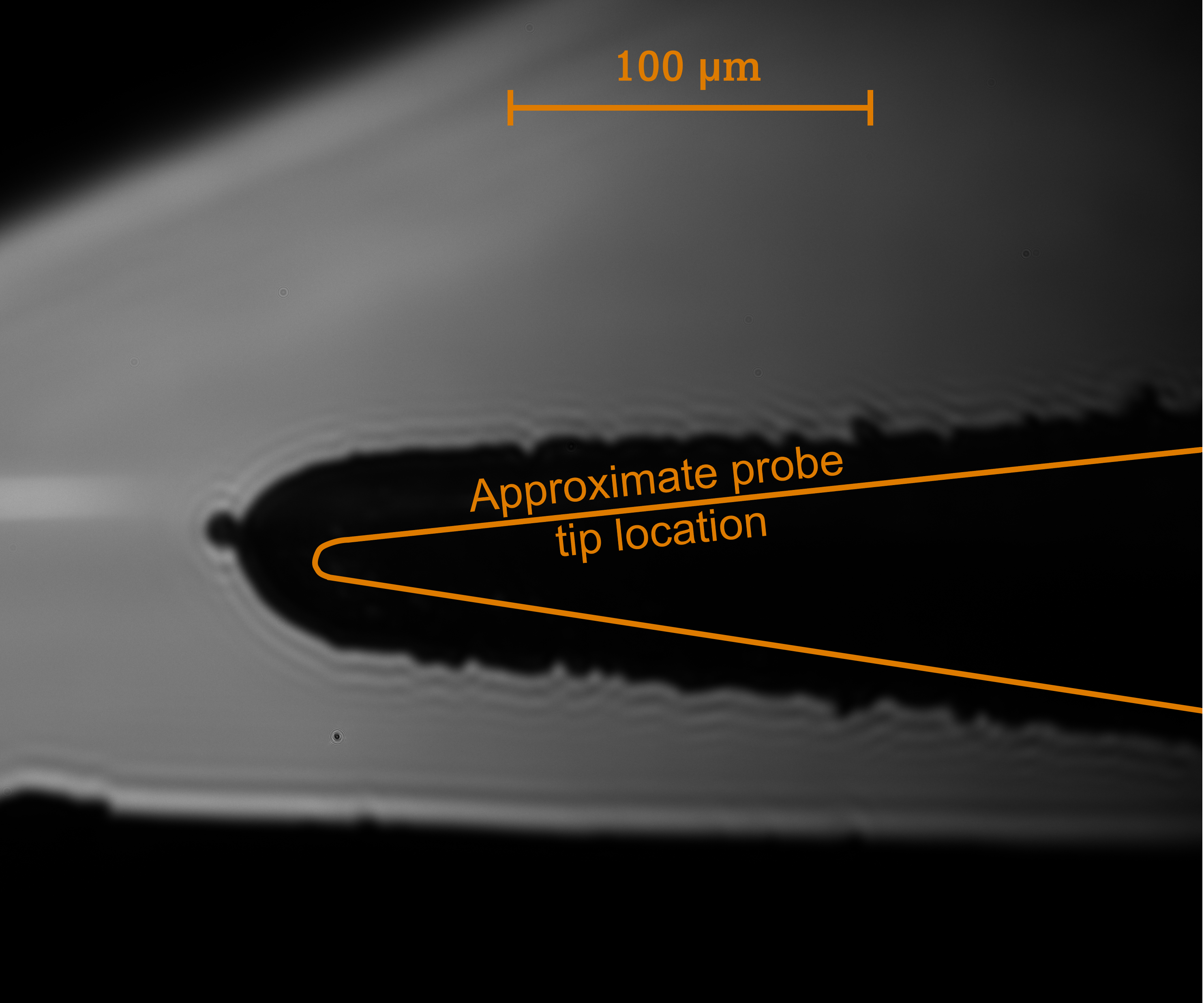}
\caption{\label{fig:trapping} Once a particle has been pressed into the camphor, the probe tip is placed into the trapping region using a micropositioner. The particle shown fell off less than a minute after this image was taken (see Figure~\ref{fig:trapped}). For this step it is vital that the camera being used is focused on the trapping region at this point; otherwise it is likely that the particle will fall off without ever entering the trap.}
\end{figure}
\end{enumerate}

\begin{figure}[ht!] 
\centering \includegraphics[width=1.0\columnwidth]{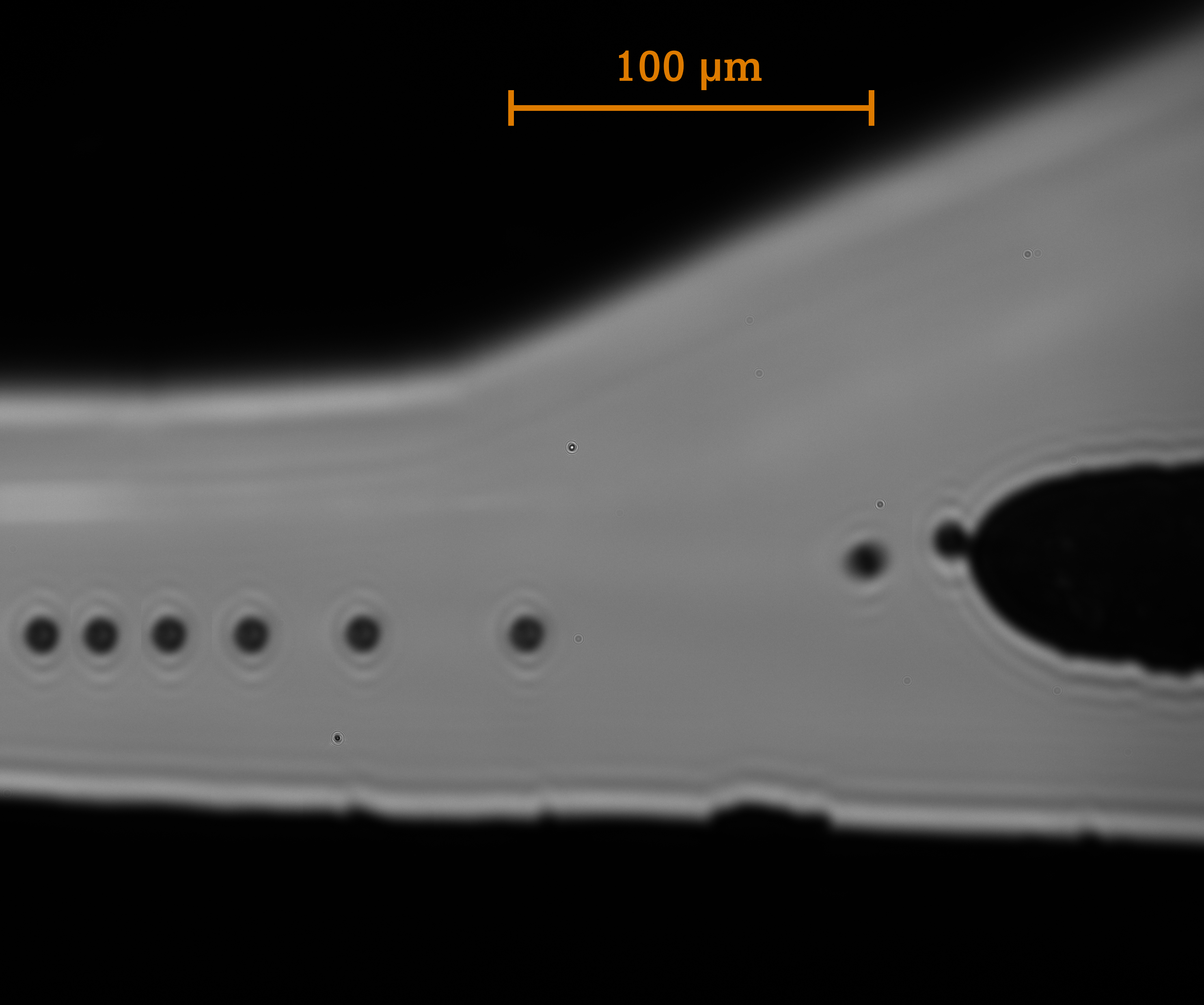}
\caption{\label{fig:trapped} Several images overlaid as the particle is released from the camphor and enters the trapping region. Images are 20 milliseconds apart. \added{Note that the motion of the particle shown in these images is determined not only by gravity, but by a combination of magnetic trapping forces and Coulombic forces from nearby patch charges.}}
\end{figure}

While the results presented here are for an $\SI{8}{\micro \meter}$ particle, tests to date have shown that this technique succeeds in trapping particles larger than $\SI{20}{\micro \meter}$ in diameter as well. The success rate of these tests is about 33\%, with the most common causes of failure being the particle sticking to the probe tip or falling off prior to the probe being moved to the trapping region. This method has also been successful in loading particles as large as $\SI{100}{\micro \meter}$, with some modifications to the technique. 

For loading particles larger than $\SI{20}{\micro \meter}$, we picked the particles up directly from the aluminum foil slide without transferring them to an uncoated probe tip first. Additionally, the tungsten probe was coated for 15 seconds instead of 10 seconds to form a thicker coating when loading larger particles. Finally, we used an
ultrasonic soldering horn (MBR Electronics USS-9210) to solder a wire onto the tungsten probe near its tip. This allowed us to generate Ohmic heating of the tungsten, which we found increased the rate of sublimation of the camphor to expedite the loading process.

Significantly, this SAR loading technique has succeeded in repeatedly selecting individual particles to load into a MGT at atmospheric pressure with minimal loss compared to other loading methods. The vacuum chamber was pumped down to ultra-high vacuum ($10^{-9}$ Torr) after many loading experiments had been performed, indicating little if any effect of residual camphor on the quality of the vacuum obtainable after using this loading technique.

\section{Limitations and Future Improvements}

There are two leading factors that contribute to failed sublimation-activated release loading attempts at present. 

The first is that charge on the particles can cause them to stick to the camphor as it sublimates, eventually leaving the particle stuck to the tungsten probe itself. This may make it difficult to use SAR in Paul traps because the trapping mechanism requires a charged particle. However, residual charge is likely to be less of a problem with larger particles, where gravity helps pull the particles off the tip.

The second issue is that particles may fall off the probe tip before it can be positioned correctly. This is currently what limits SAR to $\SI{8}{\micro\meter}$ or larger particles. We have found that the camphor coating sublimates at a rate of about 200-300 nanometers/second at room temperature, so even if an $\SI{8}{\micro\meter}$ particle is completely embedded in the camphor, the probe tip must be in position within approximately a minute or the particle will fall off before reaching the trap.

One way to mitigate this challenge might be to mount a thermo-electric heater/cooler onto the shaft of the tungsten probe to control the temperature of the tungsten. This way the tip could be kept cold (slowing the sublimation process) until the particle is ready to be trapped, at which point the probe could be heated to rapidly sublimate the camphor. 

\added{The dependence of camphor's vapor pressure on temperature can be estimated by extrapolating to a region outside a set of known Antoine equation parameters for camphor\cite{Wilde1937Uber,NISTcamphor}. At \SI{27}{\celsius}, the vapor pressure of camphor is around 90~mTorr. By comparison, small thermoelectric coolers are capable of reaching a temperature of about \SI{-40}{\celsius} with their hot side at room temperature, which yields a vapor pressure of about $1.7\times10^{-7}$ Torr. This would be expected to dramatically decrease the sublimation rate and may make it possible to adapt this technique for loading under vacuum conditions. It is also possible that cooling the probe tip during camphor deposition may affect the rate at which the heated camphor condenses onto the probe tip, potentially allowing for greater control over the uniformity of the coating.}
This is a promising direction for continued research in developing this technique.

\section{Conclusion}

This SAR loading technique allows for selective, efficient, non-contaminating, and versatile loading of micrometer-scale particles into MGTs at atmospheric pressure. Additionally, it could potentially be used with a wide variety of trapping types and geometries such as Paul traps and optical traps.

In particular, SAR loading could be a good choice for cases where the particles being loaded are irregular in shape and size, since it allows for loading a specific, pre-selected particle. It also shows the potential to be highly efficient compared to other loading techniques, which is especially helpful when the desired particles are in limited supply.

\begin{acknowledgments}
This material is based upon work supported by the National Science Foundation under Grant Nos.\ 1912083, 1950282, 2011783, and a Block Gift from the Coherent / II-VI Foundation.
\end{acknowledgments}

\section*{Data Availability Statement}

The data that support the findings of this study are available from the corresponding author upon reasonable request.

\nocite{*}
\bibliographystyle{unsrt}
\bibliography{cl_bib.bib}

\end{document}